\documentclass[pre,superscriptaddress,twocolumn,final,showpacs,showkeys,nobibnotes,titlepage,twoside,10pt]{revtex4-1}
\usepackage{amsmath}
\usepackage{graphicx}
\usepackage{amsfonts}
\usepackage{amssymb}
\usepackage{subfigure}
\usepackage{float}
\usepackage{color}

\begin{document}

\title{{Direct observation of percolation in the yielding transition of colloidal glasses
}}
\author{Antina Ghosh$^{1,2}$, Zoe Budrikis$^3$, Vijayakumar Chikkadi$^{1,2}$, Alessandro L. Sellerio$^4$, Stefano Zapperi$^{4,3}$, Peter Schall$^{1}$}

\affiliation{
$^{1}$ Institute of Physics, University of Amsterdam, Science Park 904,
1098 XH Amsterdam, The Netherlands.\\
$^{2}$ Max Planck Institute for Intelligent Systems, 70569 Stuttgart, Germany.\\
$^{3}$ ISI Foundation, Via Alassio 11C, Torino, Italy
$^4$ Center for Complexity and Biosystems, Department of Physics, University of Milano,
via Celoria 16, 20133 Milano, Italy \\
}

\begin{abstract}
When strained beyond the linear regime, soft colloidal glasses yield to steady-state plastic flow in a way that is similar to the deformation of conventional amorphous solids. Due to the much larger size of the colloidal particles with respect to the
atoms comprising an amorphous solid, colloidal glasses allow to obtain microscopic insight into the nature of the yielding transition, as we illustrate here combining experiments, atomistic simulations, and mesoscopic modeling. Our results unanimously show growing clusters of non-affine deformation percolating at yielding.  In agreement with percolation
theory, the  spanning cluster is fractal with a fractal dimension $d_f\simeq 2$, and the correlation length diverges upon approaching the critical yield strain. These results indicate that percolation of highly non-affine particles is the hallmark of the yielding transition in disordered glassy systems.
\end{abstract}


\maketitle

Soft materials like colloidal suspensions, foams and concentrated emulsions exhibit linear elastic behavior under applied strain up to a critical strain beyond which the response becomes non-linear, indicating the onset of plastic flow \cite{larson}. The microscopic origin of yielding lies in the irreversible plastic rearrangements that occur at the particle level \cite{petekidis,schall07}.
Whereas in crystals plastic deformation occurs via the motion of topological defects~\cite{defects}, in amorphous materials plasticity is associated with irreversible rearrangements of localized and highly strained zones \cite{argon,falk,schall07}.
Although rheological studies of soft glassy materials have allowed for an extensive investigation of yielding and plastic
flow at the macro scale \cite{petekidis}, their microscopic origin is still strongly debated \cite{rahmani,Jaiswal,reversibility}. Microscopic experiments so far have largely investigated particle dynamics in the steady-state regime \cite{schall07,weeks,vijay}, where plastic events are correlated by long-range quadrupolar strain fields \cite{vijay}. Such irreversible rearrangements are also observed in quiescent glasses or at small strain in the transient stages of deformation \cite{ghosh,reversibility}. What remains unclear is how these rearrangements grow and organize with increasing strain, eventually leading to yielding and plastic flow of glasses {\cite{Liu2013,percolation}}. {Experimental insight into this behavior} is of fundamental importance both for theory and for applications.

Theoretical models and simulations investigating avalanche dynamics in sheared athermal amorphous solids have focused on power-law scaling and critical behavior close to the yield point~\cite{baret2002,picard2002,dahmen,Budrikis2013,wyart,sandfeld2015,liu2016}. How far such a scaling description is valid at finite shear rates and finite temperatures is, however, a topic of active research \cite{wyart}. Experimental investigations in this direction are scarce. Recent oscillatory shear measurements of concentrated emulsions and colloidal glasses \cite{reversibility} have extended the ideas of reversible to irreversible transition (absorbing phase transition) to yielding of soft materials. These microscopic studies, which are mainly quasi two-dimensional, show that in contrast to macroscopic measurements, the microscopic signatures of yielding are indeed sharp.

In this Letter, we complement confocal microscopy experiments on three-dimensional hard-sphere colloidal glasses with atomistic simulations of metallic glasses and mesoscopic modeling, to elucidate the microscopic dynamics in the transient state across yielding. We find that highly non-affine particles form clusters that grow with strain to eventually, at a critical strain of about 10$\%$, percolate across the sample. These clusters have a fractal dimension close to 2 that remains constant with strain. Their size, as measured by the correlation length of non-affine particles, diverges upon approaching the critical strain, indicating scale-free structures. We find that the general picture is surprisingly robust across all systems studied, independent of the microscopic detail of the material, indicating that this percolation picture of yielding is much more general and applies to amorphous materials beyond colloidal glasses. However, we also find that the exponent $\nu$ governing finite-size scaling of the percolation transition is not universal, taking different values for particle-based and mesoscale models.

We use hard-sphere colloidal suspensions that are good model systems for glasses; structural relaxations slow down at particle volume fractions larger than $\phi \sim 0.58$, the colloidal glass transition \cite{pusey_megen}. Our sterically stabilized fluorescent polymethylmethacrylate (PMMA) particles have a diameter of $\sigma = 1.3 \mu m$, with a polydispersity of $7\%$ to prevent crystallization, and are suspended in a density and refractive-index matching mixture of Cycloheptyl Bromide and Cis-Decalin. The particle volume fraction is $\phi \sim 0.60$ as estimated from the centrifuged sediment, and we measured a structural relaxation time of $\tau \sim 2\times 10^4$ sec by microscopy, which is a factor of $~5\times 10^4$ larger than the Browning time $\tau_B=0.4s$. To investigate the transient deformation, we started from an equilibrated state (rejuvenation and subsequent relaxation for three hours) and applied uniform, slow shear at constant rate $\dot{\gamma} = 10^{-4} s^{-1}$, of the order of the inverse structural relaxation time. Confocal microscopy is used to image $\sim 2.5 \times 10^5$ particles in a $107 \mu m$ by $107 \mu m$ by $65\mu m$ volume, and to follow their positions in three dimensions with an accuracy of $0.03\mu m$ in the horizontal, and $0.05\mu m$ in the vertical direction~\cite{weeks_weitz}. Individual particles are tracked during a $30~min$ time interval from image stacks taken every $35~sec$, hence the experimental time increment $\delta t = 35 s$.\\

We perform molecular dynamics simulations of the compression of CuZr metallic glass using the Embedded Atom Method (EAM)~\cite{Baskes84}, as described in \cite{Mendelev2007}. Simulations are performed using the LAMMPS simulator package~\cite{Plimpton1995}, with GPU parallelization \cite{Brown11,Brown12,Brown13}. The sample is prepared starting from a Cu FCC single crystal with a lattice constant \(\lambda = 0.3610\)~nm enclosed in a simulation box with periodic boundary conditions. The alloy is generated by first transforming approximately 40\% of Cu atoms into Zr and then performing a heat treatment \cite{Sansoz20113364, nanolett5b01034} at 2300~K for 20~ps, followed by rapid quench to 10~K in 200~ps, and final relaxation at 10~K for another 20~ps.
The relaxed system is compressed along $z$ at a constant strain rate  \(2\times10^8\)~\(s^{-1}\), at \(T=10\)~K. {We confirm the results are qualitatively robust upon varying the strain rate}. Temperature and pressure are controlled using a standard Nos\'{e}-Hoover thermostat and a barostat  ~\cite{Martyna1994,parrinello1981polymorphic,Tuckerman2006,PhysRevB.69.134103}, with a characteristic relaxation time of 1~ps. The barostat ensures that the $xx$ and $yy$ components of the stress tensor are close to zero. %

We also simulate a fully tensorial mesoscale elasto-plastic model, similar to other models commonly employed to study yielding in amorphous media \citep{baret2002,picard2002,Budrikis2013,wyart,sandfeld2015,liu2016},
on a 3D cubic lattice of linear size $L=8,16,32,64$. Each lattice site represents an Eshelby inclusion \cite{Eshelby1957} of vanishing volume and strain $\boldsymbol{\epsilon}$. The stress on each site is the sum of uniform externally applied stress $\boldsymbol{\sigma}^{\mathrm{ext}}$ and internal stress
$\boldsymbol{\sigma}^{\mathrm{int}}$, which is given in Fourier space by
$\boldsymbol{\sigma}^{\mathrm{int}}_{ij}(\boldsymbol{q}) = \boldsymbol{G}_{ijkl}(\boldsymbol{q}) \boldsymbol{\epsilon}_{kl}(\boldsymbol{q}),$
where $\boldsymbol{G}$ is Eshelby's Green function~\cite{Eshelby1957}, subscripts refer to components $x$, $y$ and $z$ and Einstein summation is assumed. A site yields according to the Von Mises yield criterion on the deviatoric stress:
$ (3\boldsymbol{\sigma}^{\mathrm{dev}}_{ij} \boldsymbol{\sigma}^{\mathrm{dev}}_{ij})/2)^{1/2} > \sigma^y.$
The yield thresholds $\sigma^y$ are drawn for each site from a uniform distribution over $[0,1]$, and a site's yield threshold is re-drawn upon yield. The external stress is increased adiabatically slowly and is held constant during avalanches, as described previously~\cite{Budrikis2013}.

In experiments and atomistic simulations, we determine non-affine displacements of particles from the affine transformation of nearest-neighbor vectors over time, as described previously~\cite{falk}. The symmetric part of the affine transformation tensor is the local strain; the remaining non-affine component $D_{na}$ has been used as a measure of plastic deformation~\cite{falk,vijay,vijay12}. We focus on particles with large non-affine displacements and define a particle as "active", if its non-affine component $D_{na}^{i} > \langle D_{na} \rangle $, where angular brackets denote the average of all articles in the system. In mesoscale simulations, active sites are just the sites where plastic slip takes place.

\begin{figure*} [!]
\centering
\includegraphics[width=0.7\textwidth]{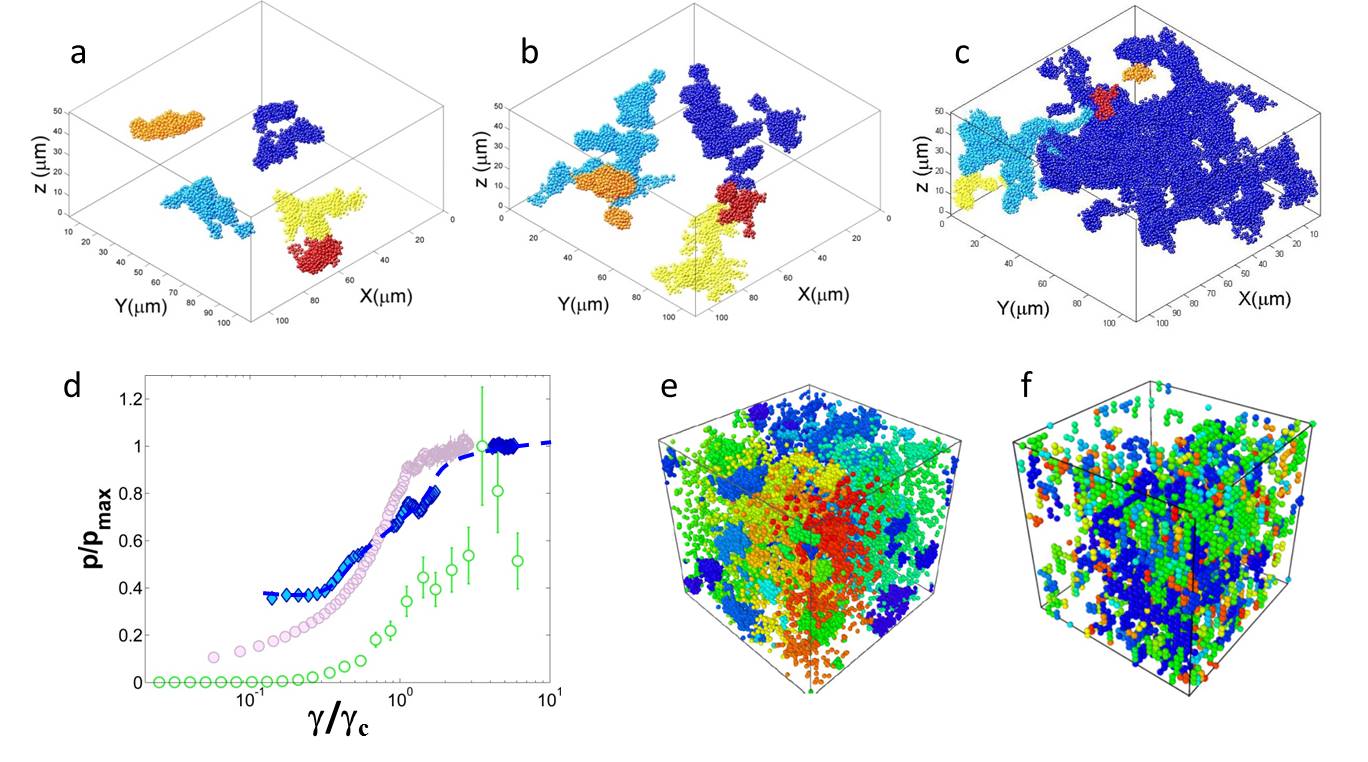}
\caption{{\bf Observation of the microscopic yielding transition} (a-c) Evolution of highly non-affine clusters in experiments at strains 2.1, 4.9, and 10.1$\%$. (d) Evolution of fraction $p$ {normalized by the maximum fraction $p_{max}$} of active sites with strain in experiments (diamonds), atomistic (pink circles) and mesoscale simulations (green circles). (e) Clusters of highly non-affine particles in atomistic simulation at $2\%$ strain. (f) Clusters of active sites in mesoscale simulations at $40\%$ strain.
}
\label{fig1}
\end{figure*}

Reconstructions of the colloidal glass reveal active particles cluster in space, and the clusters grow with applied strain, as shown in Fig. \ref{fig1}(a-c). With increasing strain these 'fluid-like' clusters expand and grow in size and new clusters appear in the field of view. Subsequently, the adjacent clusters start merging and at around a critical strain $\gamma_{c} \sim 0.1$ a single largest cluster dominates the entire field of view. We plot the fraction  $p$ of active particles as a function of strain in Fig.~\ref{fig1}(d) (blue diamonds). While initially, $p$ barely changes indicating elastic-like response, with increasing strain $p$ increases steeply and eventually reaches a steady state at higher strain.
Very similar behavior is observed in the simulations: clusters of active particles grow in space, and the fraction of active particles increases steeply and eventual saturates (Fig.~\ref{fig1}(d), pink symbols). Snapshots show clusters of active particles in the later stages of the atomistic and mesoscale simulations in Figs.~\ref{fig1}(e) and (f).

\begin{figure}[h]
\centering
\includegraphics[width=1.0\columnwidth]{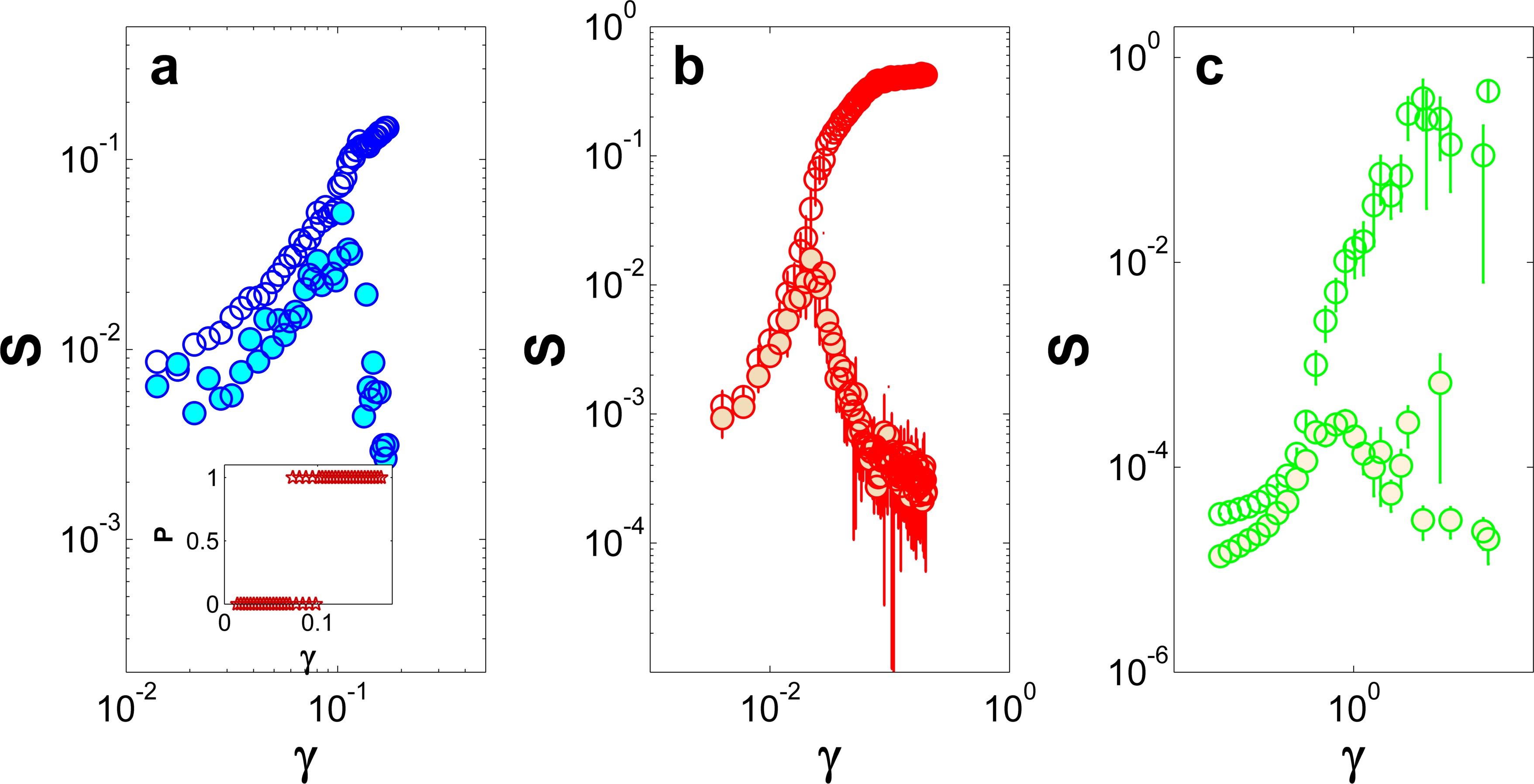}
\caption{{\bf Percolation of largest active cluster.} Size of the largest and second-largest non-affine cluster as a function of strain for experiments (a), atomistic (b) and mesoscale simulations (c). In all cases, the largest cluster grows with strain to a plateau, while the second largest cluster decays after some critical strain. The transition defines the critical strain $\gamma_c$. Inset in (a) shows percolation of largest cluster (P=1) versus strain in experiment.}
\label{fig2}
\end{figure}

We highlight the growth of the largest cluster by following the number of particles $S$ in the clusters as a function of strain. Fig.~\ref{fig2}(a) shows the evolution of the largest and second largest cluster. Both increase initially with strain, but at some critical strain $\gamma_c$ the largest cluster takes over: the second largest cluster stops growing and shrinks, while the largest cluster continues to grow, until its size eventually saturates. We use the cross-over strain $\gamma_{c}$ to define the microscopic yielding transition of the material. This critical strain is approximately $9-10\%$ in our colloidal glass, comparing well with the macroscopic yielding transition in rheological studies of hard-sphere glasses~\cite{pham,rahmani}, where yield strains of around $10\%$ are found for $\phi \sim 0.60$. A remarkably similar scenario is observed in the simulations, both atomistic and mesoscopic. Both cluster sizes initially increase, while at a critical strain, the largest cluster takes over, and the second largest cluster shrinks. This largest cluster tends to span the entire field of view, as shown for the experiments in Fig.~\ref{fig2}(a) inset, where we plot the occurrence of percolation as a function of strain.

\begin{figure}[h]
\centering
\begin{minipage}{7.0cm}
\includegraphics[width=0.9\columnwidth]{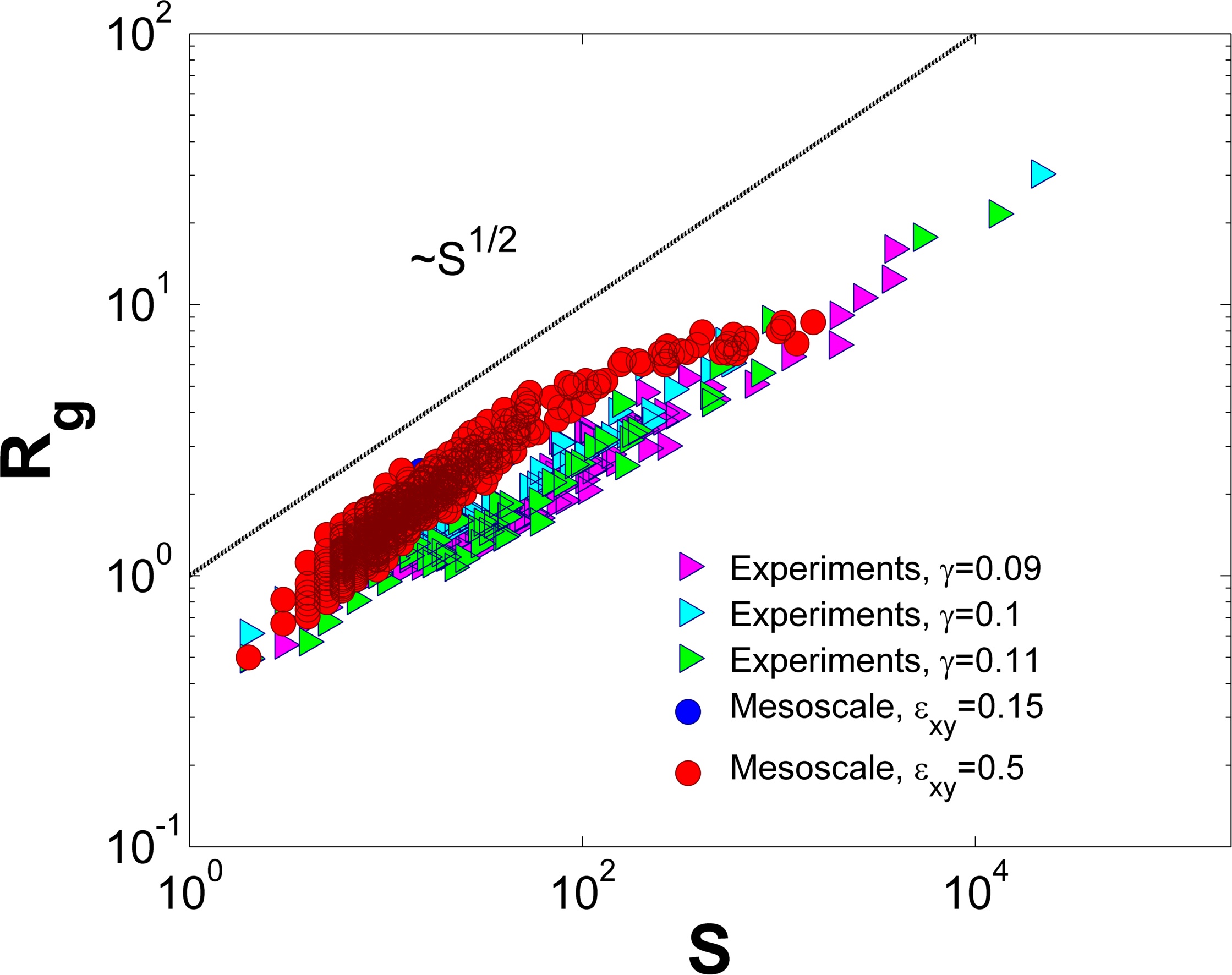}
\end{minipage}

\caption{{\bf Fractal cluster structure.} Scaling of the radius of gyration $R_g$ with size $S$ of highly non-affine clusters in experiments (triangles) and mesoscale simulations (dots) at various strain values.  The exponent $1/2$ indicates a cluster fractal dimension of 2.
}
\label{fig3}
\end{figure}

\begin{figure*}[!]
\begin{tabular}{llll}
\includegraphics[width=0.23\textwidth]{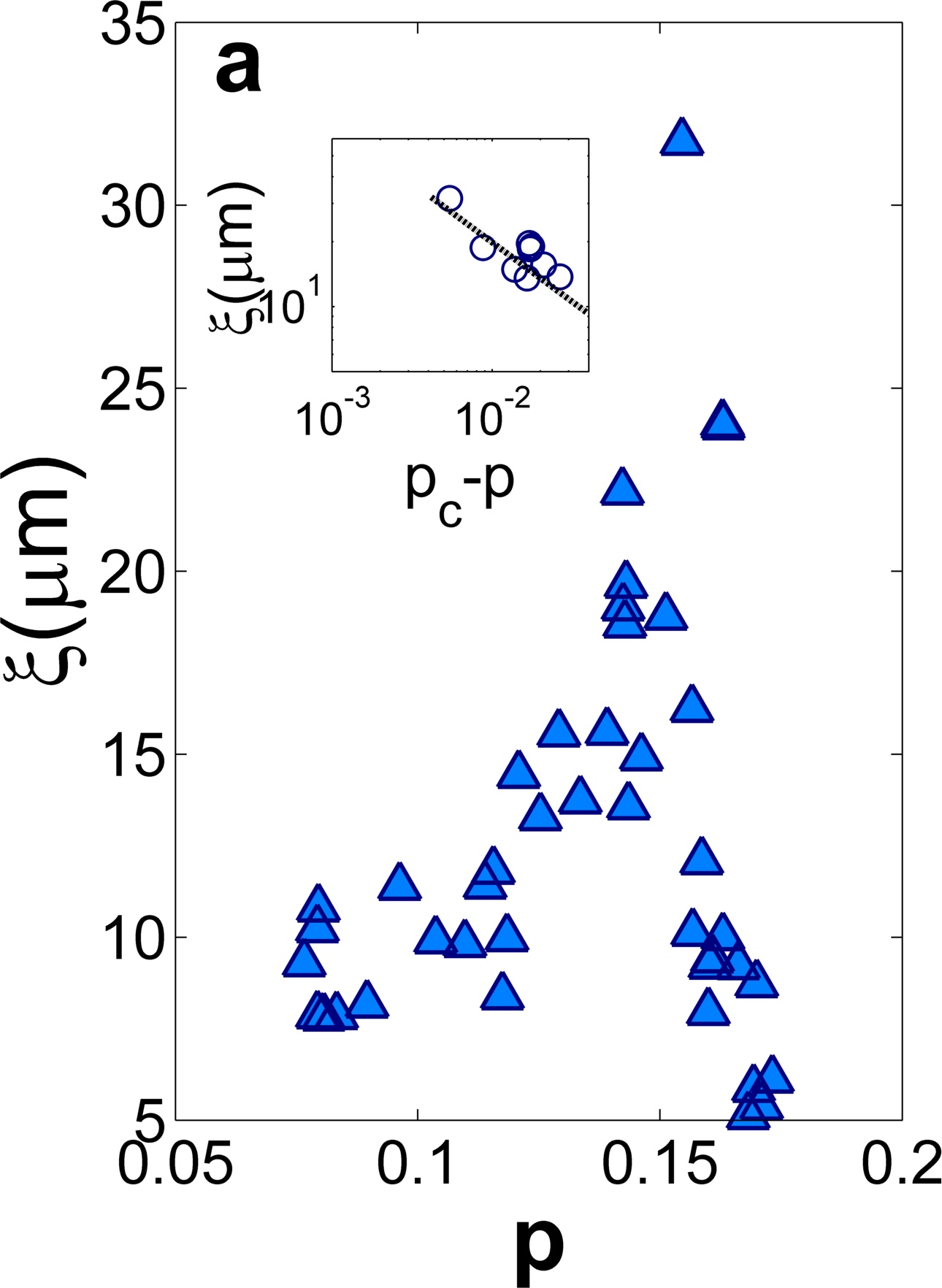}&
\includegraphics[width=0.26\textwidth]{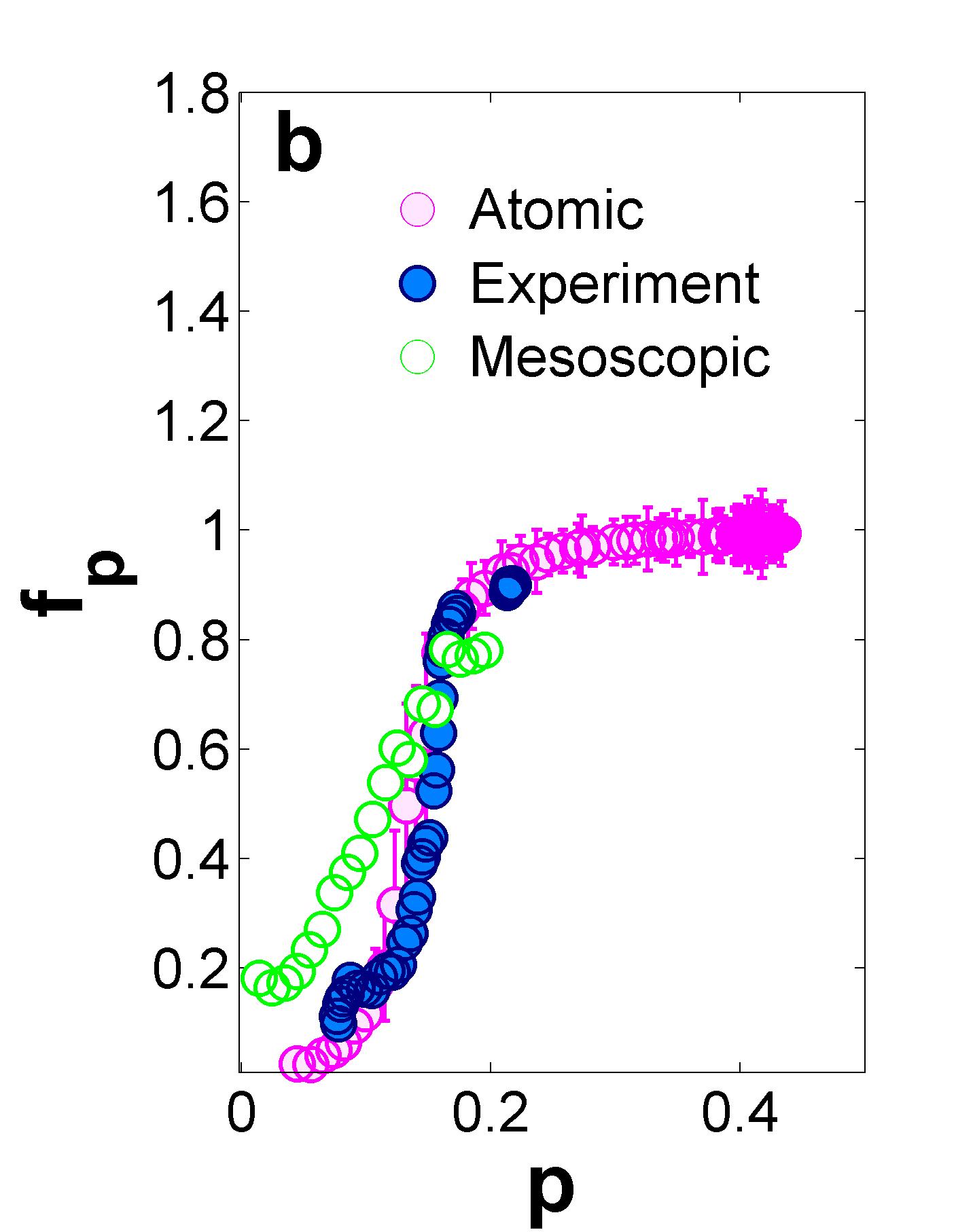}&
\includegraphics[width=0.23\textwidth]{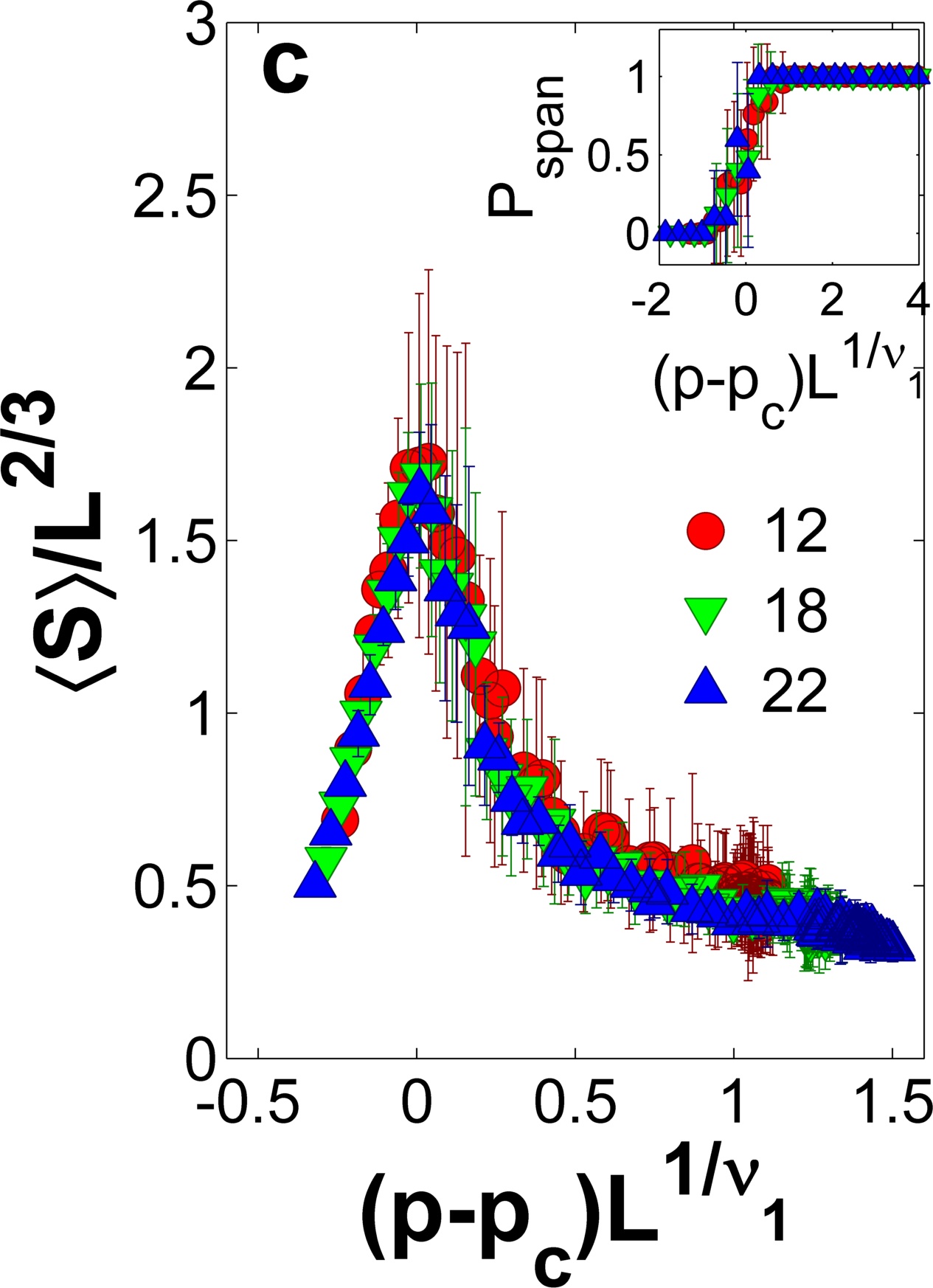}&
\includegraphics[width=0.23\textwidth]{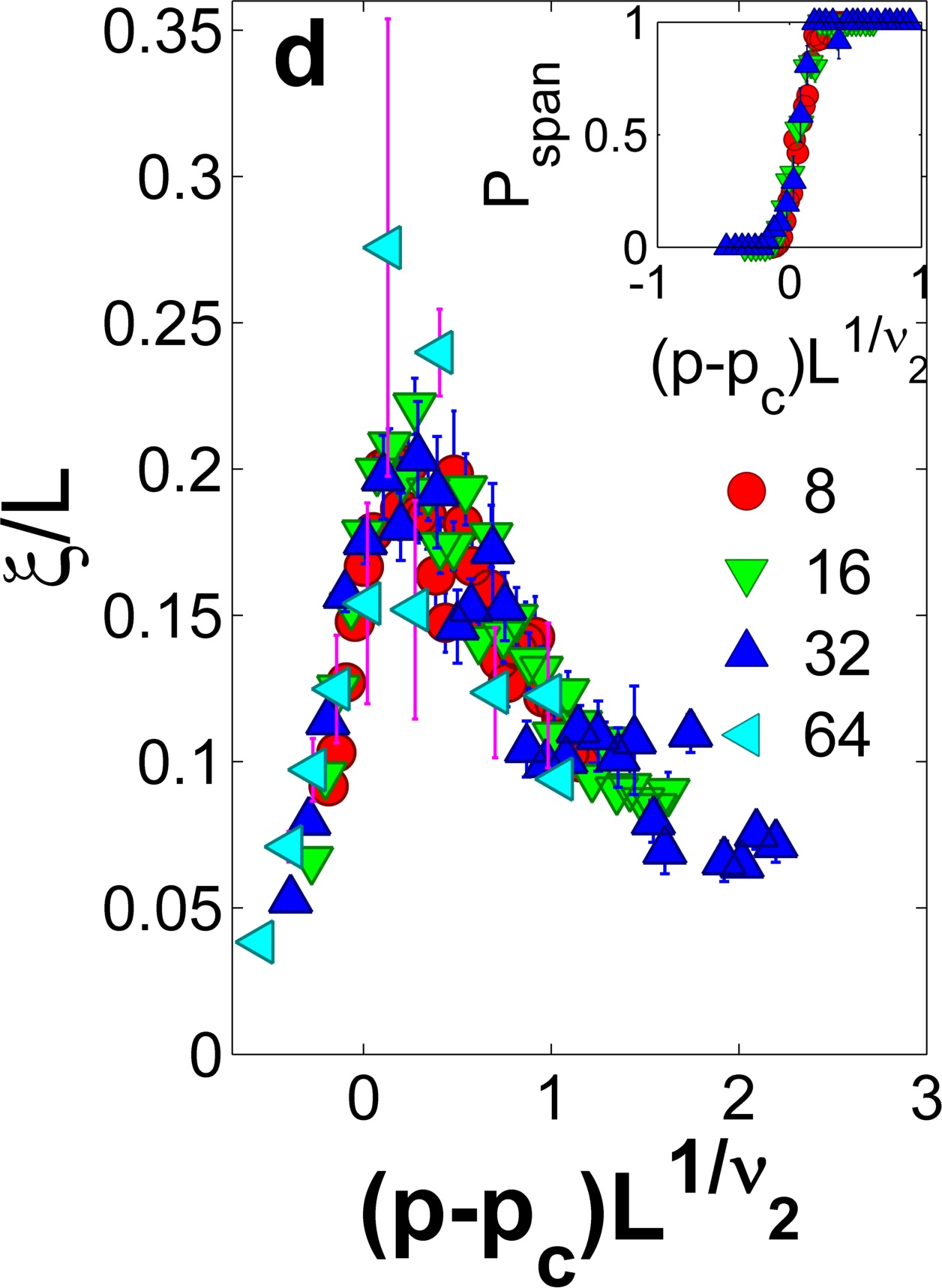}\\
\end{tabular}
\caption{{\bf Cluster divergence and percolation} \label{fig4}
(a) Cluster correlation length $\xi$ as a function of fraction $p$ of active colloidal particles. Inset: Scaling of $\xi$ with distance from the critical fraction $p_c \sim 0.16$. The correlation length diverges according to $\xi \sim (p_{c} - p)^{-\nu}$ with $\nu \sim 0.75$. (b) Fraction $f_{p}$ of particles in the largest cluster as a function of total fraction of active particles. For the colloidal glass, the critical fraction where the transition from non percolated to a percolated state happens is again $p_{c} \approx 0.16~(f_p\simeq 1/2$). (c) Finite-size scaling collapse of the mean cluster size in atomistic simulations. Inset: probability $P_{span}$ of finding a spanning cluster as a function of rescaled distance $p_{c} - p$. The critical fraction $p_c=0.113 \pm 0.001$ and exponent $\nu=\nu_1=0.85 \pm 0.1$ are obtained through a joint fit over data sets for system sizes $L=12,18,22$~nm of the probability $P_{\mathrm{span}}(p)$ of finding a spanning cluster, using the size-dependent sigmoid $1/(1+\exp(-k(p-p_c)L^{1/\nu}))^a$, as shown in the inset. Here $k=3.9\pm1.3$ and $a=0.82\pm0.07$ are sigmoid shape parameters, and all fitted parameters are shared between all data sets.
(d) Similar collapse for mesoscale simulations where the fit yields $\nu=\nu_2=2.1\pm 0.03$.
}
\end{figure*}

We find that the clusters have fractal shape. To show this we compute the cluster radius of gyration $R_{g}^{2} = 1/2 (\sum_{ij} (r_{i} - r_{j})^{2} / S^{2})$ \cite{stauffer}, as a function of cluster size $\it{S}$, which we plot in Fig.~\ref{fig3}. The radius of gyration scales with cluster size $S$ as $R_{g} \sim S^{1/d_{f}}$ with $d_{f} \sim 2$, indicating that the clusters have a fractal dimension  $d_{f} \sim 2$. We find that this scaling is robust and independent of the applied strain. This fractal structure is in line with the hierarchical organization of plasticity observed in the steady-state flow after yielding~\cite{vijay}, and indicates a near-critical state of the system. \\

To investigate the growth of fractal clusters upon approaching $\gamma_c$, we compute the characteristic length scale of non-affine regions. We determine the correlation length of clusters of non-affine particles using $\xi^2 = 2 \sum_i R_{gi}^2 S_i^2 /\sum_i S_i^2 $ where $R_{gi}$ is the radius of gyration for cluster size $S_i$ \cite{stauffer}. This correlation length increases with the increasing fraction $p$ of active particles and diverges near a critical fraction $p_c \sim 0.16$, at the critical strain $\gamma_c$, as shown in  Fig.~\ref{fig4}(a). Around this strain, we measure a correlation length of $\xi \sim 32 \mu m$, of the order of the thickness of the sheared colloidal layer of $\sim 50 \mu m$.  {This growth of correlation length is in line with the growing correlation time scale observed in oscillatory yielding experiments~\cite{reversibility}}. Furthermore, by plotting the correlation length as a function of the distance $p_c - p$ to the critical fraction (Fig.~\ref{fig4}(a), inset), we find that the correlation length grows with a power law $\xi \sim (p_c - p)^{-\nu}$ upon approaching the critical fraction $p_{c}$. Here, $\nu \sim 0.75$. This exponent appears close to that predicted for percolation in three-dimensional continuum percolation models~\cite{Continuum_percolation, stauffer}.

The emerging picture is thus that regions of highly non-affine, fluid-like particles grow and eventually, at the yielding transition, percolate across the sample. To test this idea in more detail, we apply concepts from percolation theory and follow the size of the largest cluster as a function of the total number of active particles. We plot the fraction $f_{p}$ of particles in the largest cluster as a function of the total fraction of active particles in Fig.~\ref{fig4}(b). This fraction increases sharply at $p_{c}$, indicating that the largest cluster abruptly takes over and absorbs all active particles. This scenario is indeed characteristic for percolation: the fluid-like particles that percolate at yielding produce a fluidized network that sustains the steady-state flow after yielding. The critical fraction of highly active particles is $p_c\sim 0.16$ at $f_{p} = 0.5$, i.e. approximately 16$\%$ of the total number of particles. The corresponding critical strain is again $\gamma_{c} \sim 9-10\%$, in good agreement with reported yield strains of colloidal glasses. We hence find that the microscopic origin of yielding is the percolation of highly non-affine particle clusters, producing a fluid-like network in a solid matrix.

Our simulations allow us to study the transition at $p_c$ in greater detail, by performing a finite-size scaling collapse of the mean cluster size $S$  as a function of active particle fraction, $p$, using the standard percolation rescaling $p\to(p-p_c)L^{1/\nu}$. Figure~\ref{fig4}(c) shows the results for atomistic simulations, where $p_c=0.113 \pm 0.001$ and exponent $\nu=0.85 \pm 0.1$.
This exponent agrees very well with the expected value for percolation in three dimensions of $\nu=0.88$ \cite{Continuum_percolation, stauffer}. Similarly, our mesoscale simulations also show a percolation-like transition at $p_c=0.085\pm0.005$, but with exponent $\nu = 2.1\pm0.03$, as shown by the excellent fits of $P_{\mathrm{span}}$ and the data collapse of $\xi$ in Fig.~\ref{fig4}(d). The different scaling exponent appears to be a particular feature of the mesoscopic model that is at odds with atomistic simulations and experiments.
This may suggest that models including only linear elasticity and quenched disorder \cite{wyart,Budrikis2013}, might
be too simple to recapitulate the detailed scaling features of the percolation transition associated with amorphous yielding.

To summarize, we have used experiments on colloidal glasses and atomistic and mesoscale simulations to show that the microscopic yielding of glasses originates from the percolation of non-affine, plastic regions. Non-affine particles form clusters that grow with applied strain and eventually merge. At some critical fraction of non-affine particles, the largest cluster abruptly takes over and absorbs all other non-affine particles to produce a percolated network. The non-affine clusters themselves have fractal shape, and upon approaching the yielding transition, their size diverges in a critical fashion. The robust fractal dimension and its identical value in colloidal experiments and simulations points towards a universal critical transition at the yielding of glasses.

The general percolation phenomenology we uncover here is robust and appears regardless of the microscopic detail of the system studied.
We have reported similar results in experiments on colloidal glasses, where particles have a micrometer size, in simulations of metallic glasses where particles are at atoms, and in mesoscale simulations where particles are not even present. This suggests a common scenario ruled by the interplay between structural disorder and elasticity, which are the two common ingredients of the systems we study. However, it is less clear that the phenomena are strictly universal in terms of critical exponents and scaling functions. While in colloidal and metallic glasses, clusters are described by three-dimensional conventional percolation scaling, our mesoscale model yields a different exponent $\nu$. This result raises interesting questions on the most appropriate coarse-grained description of the
yielding of amorphous solids.

\section{Acknowledgements}
V.C. and P.S. acknowledge support by VIDI and VICI fellowships from NWO (Netherlands Organization for Scientific Research). A.L.S., Z.B. and S.Z. are supported by the European Research Council (ERC) Advanced Grant 291002 SIZEFFECTS.

\end{document}